\input epsf.sty
\documentstyle[aps,prd]{revtex}

\def\gtsima{$\; \buildrel > \over \sim \;$}
\def\ltsima{$\; \buildrel < \over \sim \;$}
\def\prosima{$\; \buildrel \propto \over \sim \;$}
\def\gsim{\lower.5ex\hbox{\gtsima}}
\def\lsim{\lower.5ex\hbox{\ltsima}}
\def\simgt{\lower.5ex\hbox{\gtsima}}
\def\simlt{\lower.5ex\hbox{\ltsima}}
\def\simpr{\lower.5ex\hbox{\prosima}}

\tighten
\begin{document}

\title{Head--on Collision of Two Unequal Mass Black Holes}
\author{Peter Anninos${}^{(1)}$ and Steven Brandt${}^{(2)}$}
\address{
${}^{(1)}$ National Center for Supercomputing Applications,
Beckman Institute, 405 N. Mathews Avenue, Urbana, Illinois, 61801 \\
${}^{(2)}$ Max--Planck--Institut f\"ur Gravitationsphysik
(Albert--Einstein--Institut), Schlaatzweg 1, 14473 Potsdam, Germany
}

\date{\today}
\maketitle

\begin{abstract}
We present results from the first fully nonlinear numerical calculations
of the head--on collision of two unequal mass black holes.
Selected waveforms of the most dominant $\ell=2$, 3 and 4
quasinormal modes are shown, as are the total radiated energies
and recoil velocities for a range of mass ratios and
initial separations.
Our results validate the close and
distant separation limit perturbation studies, and suggest
that the head--on collision scenario is not likely to
produce an astrophysically significant recoil effect.
\end{abstract}


The collision of two black holes is expected to be one of the
most interesting and important astrophysical sources of gravitational 
radiation. The strong bursts of radiation emitted from the
interaction should be detectable by the next generation 
gravitational wave observatories,
and will provide explicit evidence for the existence of black holes,
assuming the characteristic signatures of the holes are observed.
To compliment the anticipated observations, several calculations
of binary black hole interactions have been performed in recent
years, inspired in part by a broader Grand Challenge
effort to supply waveform templates that are essential
for the analysis of observational data. These calculations
have been based on axisymmetric numerical relativity computations 
of the head--on collision of two initially time--symmetric and boosted
black holes \cite{Anninos1,Baker},
as well as perturbation studies \cite{Baker,Anninos2}
which confirm and elucidate the numerical results in the 
close limit approximation (CLAP).
However, the fully nonlinear numerical studies performed until now
have been restricted to the simplified case of equal mass black hole
systems. To extend these calculations, we have recently introduced
a new class of body--fitting coordinate systems 
(class I in the notation of Ref. \cite{Anninos3}) that
overcomes the coordinate singularity problems in previous work.
In this paper we utilize these new coordinates to carry out the
first numerical studies of
axisymmetric head--on collisions of unequal mass black holes.
In addition to expanding the catalog of
computed energies and waveforms, numerical investigations of these systems
can address whether gravitational radiation
from the collisions is likely to carry an astrophysically significant
component of linear momentum, and whether substantial recoil velocities
can be imparted to the final black hole.

The initial data used in our simulations are 
time--symmetric analytic solutions that generalize the equal
mass Misner \cite{Misner} data adopted in previous numerical 
calculations \cite{Anninos1}. Applying the isometry operation systematically
across each of the throats for the infinite image poles, the conformal
factor can be written in cylindrical coordinates as
\begin{equation}
\Psi = 1 + \sum_{n=1}^{\infty}\left[
           \frac{{\rm csch}{\mu_{n,1}}}
                 {\sqrt{\rho^2+\left(z-{\rm coth}{\mu_{n,1}}\right)^2}}
         +\frac{{\rm csch}{\mu_{n,2}}}
                {\sqrt{\rho^2+\left(z+{\rm coth}{\mu_{n,2}}\right)^2}} 
          \right] , 
\label{eqn:confact}
\end{equation}
where we have defined $\mu_{n,1} = \mu_{n,2} = n\mu_0$ for even $n$,
$\mu_{n,1} = \mu_1+(n-1)\mu_0$ and
$\mu_{n,2} = \mu_2+(n-1)\mu_0$ for odd $n$, 
and  $\mu_0 = (\mu_1 + \mu_2)/2$.
The initial data is thus uniquely specified
by two parameters, $\mu_1$ and $\mu_2$, which determine
the black hole masses and separations, analogous
to the single Misner parameter in the equal mass case.
The physical content of the initial data is characterized by the
total (ADM) mass of the spacetime derived as the asymptotic leading
order term in $M_{ADM}=-2 r^2 \partial_r \Psi$
\begin{equation}
M_{ADM} = 2 \sum_{n=1}^\infty \left[ {\rm csch}\left(\mu_{n,1}\right)
          +{\rm csch}\left(\mu_{n,2}\right) \right] ,
\label{eqn:admmass}
\end{equation}
the proper distance separating the black hole throats 
along the ($\rho=0$) axis
\begin{equation}
L = 2 + \frac{2\mu_2}{\sinh\mu_2} + 2\sum_{n=1}^{\infty}\left[
    \frac{4n\mu_0}{\sinh 2n\mu_0}
  + \frac{2n\mu_0-\mu_2}{\sinh(2n\mu_0-\mu_2)}
  + \frac{2n\mu_0+\mu_2}{\sinh(2n\mu_0+\mu_2)} \right] ,
\end{equation}
and the bare masses of the two holes as suggested by 
Lindquist \cite{Lindquist,Andrade}
\begin{eqnarray}
m_1 &=& 2\sum_{n=1}^{\infty} \left[\frac{2n}{\sinh 2n\mu_0}
     + \frac{n}{\sinh(2n\mu_0-\mu_2)} +\frac{n}{\sinh(2n\mu_0+\mu_2)}\right] , \\
m_2 &=& 2\sum_{n=1}^{\infty} \left[\frac{2n}{\sinh 2n\mu_0}
     + \frac{n}{\sinh(2n\mu_0-\mu_1)} +\frac{n}{\sinh(2n\mu_0+\mu_1)}\right] .
\end{eqnarray}

The numerical solutions are parameterized by the ratio of bare
masses $M_R=m_1/m_2$ with 
the convention $m_1 \le m_2$, and the proper separation
between the throats $L/M$ normalized to the maximum of the
two masses. Here we simply define
$M = M_{ADM}/(1+M_R)$ for consistency with previous work.
The code is applied to collide black holes with initial separations up
to $\sim 10 M$ with mass ratios $M_R = 1$, 0.75, 0.5 and 0.25. 
These choices are determined mainly by stability issues, which
require $(\mu_1,~\mu_2) \simlt 3.0$ as the upper bound \cite{Anninos3},
and accuracy issues which require $(\mu_1,~\mu_2) \simgt 0.5$ 
to resolve the very weak higher order ($\ell \ge 3$) multipole waveforms,
although an even greater range of parameters can be evolved accurately
in the stronger $\ell=2$ signals.
For each of the cases presented, the
code is stable enough to maintain accurate solutions to well past
$100 M_{ADM}$, which is more than enough time to monitor the
complete wave signals with current parameters.
Although results are presented for $M_R \ge 0.25$, we
expect that this range of parameters will produce maximal
or near maximal recoil velocities, a conclusion supported by
the CLAP \cite{Andrade} and Fitchett \cite{Fitchett1} calculations
which predict maximum recoil at $M_R\sim 0.25$ and 0.38
respectively.

The method we use to compute waveforms is based on the
technique developed by Abrahams and Evans \cite{Abrahams1} (and
applied in Ref. \cite{Abrahams2}) to extract
the Regge--Wheeler perturbation functions and to construct the
gauge invariant Zerilli function $\psi$.
Fig. \ref{fig:waveforms} shows a characteristic sample of generated
waveforms for the particular case of 
$\mu_1 = 2.5$ and $\mu_2 = 2$ corresponding to
$M_R=0.66$ and $L/M = 7.45$.
Also shown are the quasinormal mode fits from perturbation theory using 
a linear combination of the fundamental and first overtone
for the final black hole of mass $M_{ADM}$.
The fits nicely match both the frequency and damping rate
in the late--stage ringing phase.
In Fig. \ref{fig:energy} we show the total energy radiated in the
dominant $\ell=2$ mode as a function of initial separation distance
for different mass ratios.
For comparison, we note that the perturbation results of
Davis et al. (DRPP) \cite{Davis}, when rescaled by the reduced
mass, yields 
$E/M_{ADM}\approx 0.0104 \mu^2/M_{ADM}^2 \approx 0.0104 M_R^2/(1+M_R)^3$ 
for the radiated energy in the large separation limit,
where $\mu$ is the reduced mass.
Our numerical results are in remarkable agreement with this
prediction, which suggests that the radiated energy
(when normalized to the ADM mass) decreases as $M_R^2$ and
yields maximum energy emissions of (3, 7, 10 and 13)$\times 10^{-4}$
for the $M_R =$ (0.25, 0.5, 0.75 and 1.0) cases respectively.
In the opposite close separation limit, our results are also
in excellent agreement with the CLAP calculations of
Andrade and Price \cite{Andrade}, and the particle
limit ($M_R \rightarrow 0$) collisions 
from finite initial separations based on the Green's function
integration by Lousto and Price\cite{Lousto}.
This last curve appropriately bounds the
sequence of data points from our numerical relativity calculations.

As indicated by Fig. \ref{fig:waveforms}, the quasinormal modes
are strongly excited, and the black hole oscillations are well described by
essentially the quasinormal modes (and particularly the $\ell=2$ mode)
of the final black hole despite the fact
that the systems considered are not small perturbations.
In this regard, the collision of unequal mass black holes is similar
to that of equal mass systems, except now the breaking of
equatorial symmetry allows for odd mode multipole components.
The mixing of adjacent multipole modes gives
rise to a nonvanishing flux of linear momentum along the 
$z$--axis which can be evaluated using the
Landau--Lifshitz pseudotensor \cite{Landau,Cunningham}
\begin{equation}
\frac{d P^z}{dt} = \frac{1}{32 r^2} \int_0^\pi
                   \left(\frac{\partial g_{\theta\theta}}{\partial t}
                       - \frac{\partial g_{\phi\phi}}{\partial t}
                         \frac{1}{\sin^2\theta} \right)^2 
                  \cos\theta\sin\theta d\theta .
\label{eqn:landau}
\end{equation}
As an additional check we have also computed the momentum flux
from consecutive Zerilli components \cite{Andrade}
\begin{equation}
\frac{dP^z}{dt} = \frac{1}{16\pi} \sum_{l=2}^{\infty}
                  \sqrt{\frac{(\ell-1)(\ell+3)}{(2\ell+1)(2\ell+3)}}
                  \frac{d\psi_\ell}{dt}\frac{d\psi_{\ell+1}}{dt} ,
\end{equation}
and we find generally good agreement 
(typically 10 - 30\% relative differences) between the two calculations.
As a result of the momentum emission from 
odd/even mode mixing, the final black hole
will acquire a recoil velocity
$v^z = - M_{ADM}^{-1} \int (dP^z/dt) dt$,
opposite in direction to the momentum flux of the waves.

The radiation of momentum by gravitational waves 
is of interest in work relating to
active galactic nuclei, quasars and even archetypical galaxies
since highly dynamic black holes and coalescing
binary black hole systems may abound in galactic discs as well as in
the centers of galactic nuclei. Because the efficiency of momentum
radiation emission is not known precisely, the stability of
these systems remains in question due to the
generation of a recoil velocity in the affected black holes,
which, for sufficiently asymmetric configurations, may be large
enough to break black holes free from the host galaxy.
If so, gravitational radiation effects will
have considerable observable consequences for astrophysics and
cosmology (such as the depletion of black holes
from host galaxies, the disruption of active galactic core energetics,
and the introduction of black holes and stellar material into the IGM)
since the processes generating the momentum emission,
i.e., stellar core collapse and binary mergers,
are likely to be fairly common.

Although fully nonlinear numerical simulations of
asymmetric black hole systems have not been
performed to now, several quasi--Newtonian and perturbation
calculations have been carried out over the past two decades
to estimate the magnitude of this recoil effect.
For example, Bekenstein \cite{Bekenstein} derived an upper bound
of $v_r \le 300$ km/sec for the recoil velocity
from nonspherical stellar core collapse;
Moncrief \cite{Moncrief} found $v_r = 25$ km/sec for
small non--spherical distortions 
in Oppenheimer--Snyder collapse models; and
Nakamura and Haugan (NH) \cite{Nakamura} 
computed $v_r = |\Delta p|/M \sim 262 (m_1/M)^2~\mbox{km/sec}$
for a test particle of mass $m_1$ plunging from rest at
infinity into a Schwarzschild black hole of mass $M$, with $m_1 \ll M$. 
Extrapolating the NH result to the comparable mass limit
($m_1 \sim M$), we replace the mass dependency with
Fitchett's \cite{Fitchett1} scaling formula,
$(m_1/M)^2 \rightarrow f(M_R) = M_R^2 (1-M_R) (1+M_R)^{-5}$,
which has the desired perturbation and equal mass limits.
Considering the maximal value of $f = 0.01789$
at $M_R = 0.38$, the perturbation result
predicts a maximum recoil velocity of about 5 km/sec
for the head--on collision scenario.

However, numerical
calculations are needed to adequately resolve the equal or near--equal
mass cases, and to properly model nonlinear effects from
internal dynamics such as tidal heating, and the re--absorption
and beaming of gravitational waves.
We plot in Fig. \ref{fig:momentum} the absolute value of the
recoil velocities found in our simulations by evaluating the integrated 
momentum flux from the
Landau--Lifshitz pseudotensor (\ref{eqn:landau}) across a
2--surface of radius $15 M_{ADM}$. Despite the fact that we
are unable to accurately evolve black hole collisions
with initial separations greater than about $10 M$, 
Fig. \ref{fig:momentum} does suggest a maximum recoil on the order
of $v_r \sim 10 - 20$ km/sec, in rough agreement with the
various perturbation and quasi--Newtonian estimates.
At sufficiently small initial separations, our
results also agree nicely with CLAP, especially when considering
the smallness of this effect and the 
estimates of numerical errors based on
resolution studies (less than 10\% variation between 
fixed detectors in the 200 and 300
radial zone evolutions) and extractions of the 
momentum across differently positioned detectors 
(roughly 10--30\% variation
between detectors at radii 15 and 25 $M_{ADM}$).
In comparison, the escape velocity from galactic structures
with radius $R$ and total mass $M$ is of order
$\sqrt{G M/R}$, which can vary from anywhere
between about 200 km/sec for the less massive compact dwarf galaxies
to about 1000 km/sec for giant ellipticals
(though black holes in the galactic disk and far from the
core would require substantially less recoil in the direction of
galactic rotation to achieve escape velocity).
We thus conclude that the head--on collision of black holes
is not likely to produce an astrophysically significant
recoil effect.

We wish to thank Z. Andrade, R. Matzner, R. Price and L. Smarr for
many helpful discussions, and especially
ZA and RP for also making available their
CLAP data for comparison. PA is grateful to the
Albert Einstein Institute for their hospitality during which
part of this work was conducted.
The calculations were performed on the C90 at PSC and
the Origin2000 at AEI.


\figure{
\epsfysize=3.4in \epsfbox{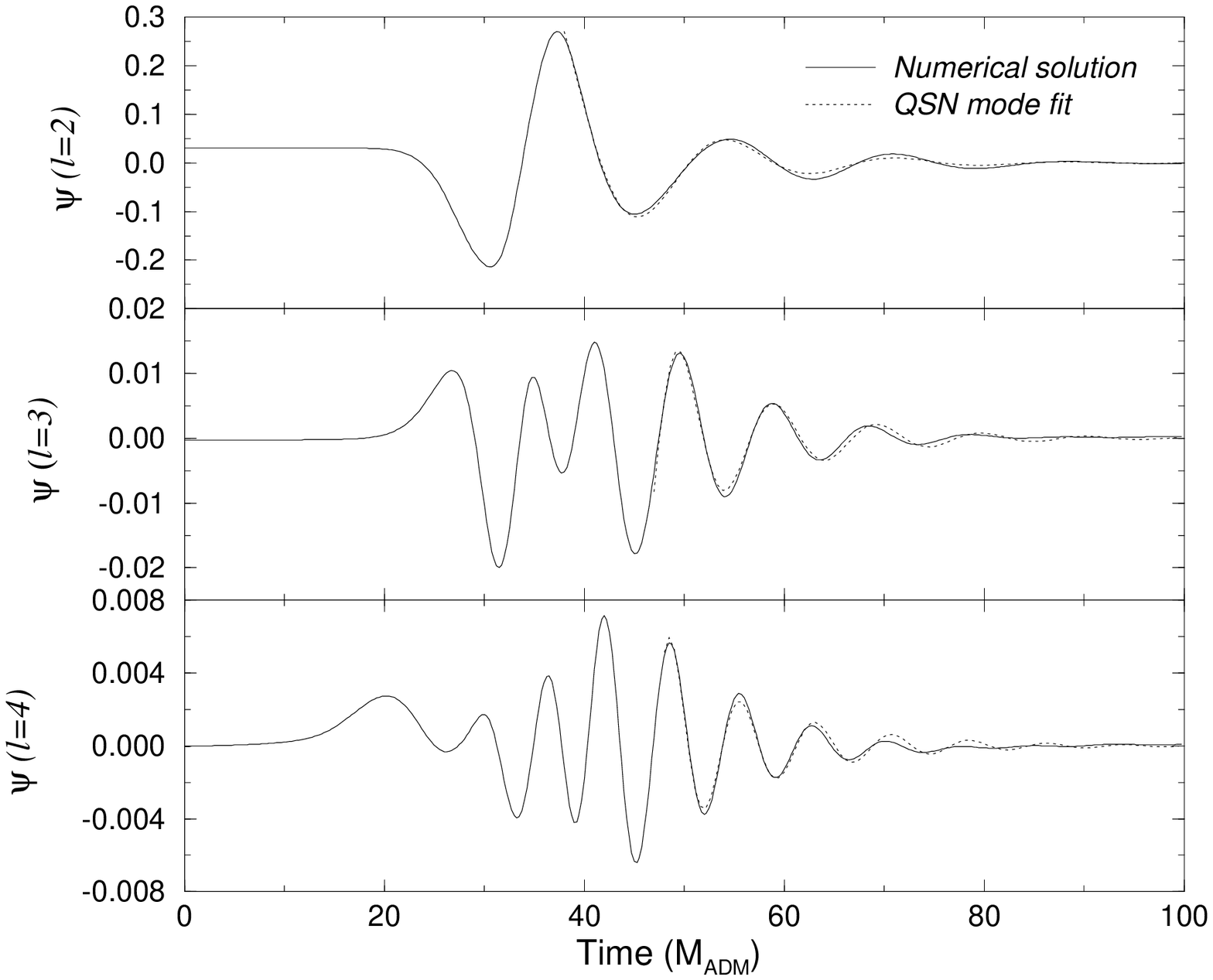}
\caption{
}
Zerilli waveforms
for the $\mu_1 = 2.5$ and $\mu_2 = 2$ case,
corresponding to a mass ratio $M_R=m_1/m_2 = 0.66$ and 
proper initial separation $L/M= 7.45$. The
$\ell=2$, 3, and 4 modes are displayed
normalized to the ADM mass, and
extracted at a radius of $20 M_{ADM}$.
Also shown with dotted lines are fits to the fundamental and
first overtone of the quasinormal modes from perturbation theory.
The simulations are run at a grid resolution
of 200$\times$70 (radial$\times$angular) zones.
\label{fig:waveforms}
}

\figure{
\epsfysize=3.4in \epsfbox{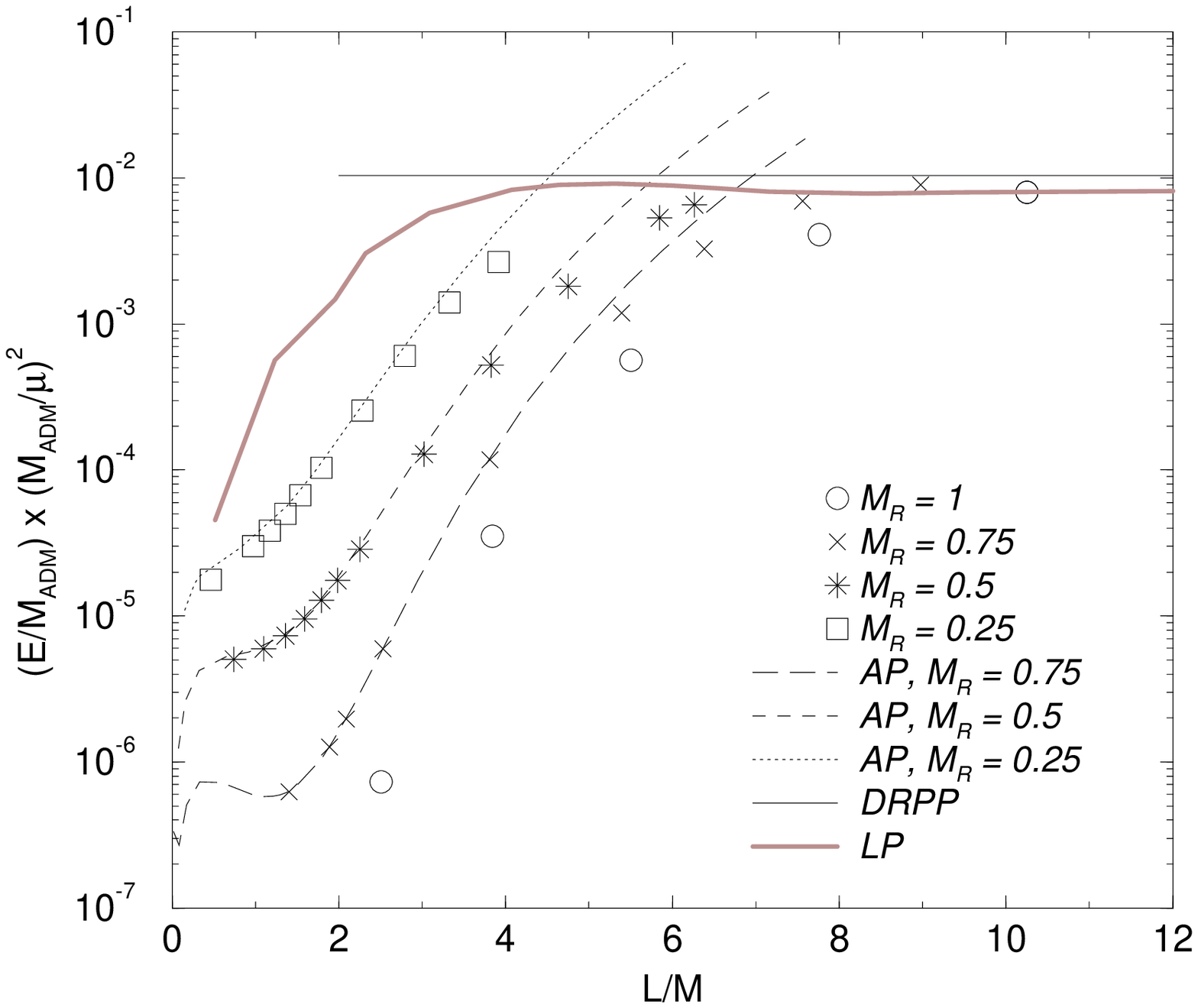}
\caption{
}
The total energy radiated in the dominant $\ell=2$ mode
from the head--on collision of two black holes
as a function of separation distance $L/M$ between the holes, where
$M=M_{ADM}/(1+M_R)$ is the approximate mass of the larger black hole,
and $M_R = m_1/m_2$ is the ratio of bare masses.
The energies are computed as the integrated flux across a
2--surface of radius $15 M_{ADM}$, and rescaled by the
ADM mass and the perturbative ($m_1/m_2 \ll 1$) behavior.
Four level curves are shown for different ratios of
bare masses $M_R= 1$, 0.75, 0.5 and 0.25, together with
the corresponding CLAP results from Andrade and Price (AP)
\cite{Andrade}, the reduced mass extension of the Davis et al.
(DRPP) \cite{Davis} perturbation calculation, and the 
Green's function integrations of Lousto and Price (LP) \cite{Lousto}.
\label{fig:energy}
}

\figure{
\epsfysize=3.4in \epsfbox{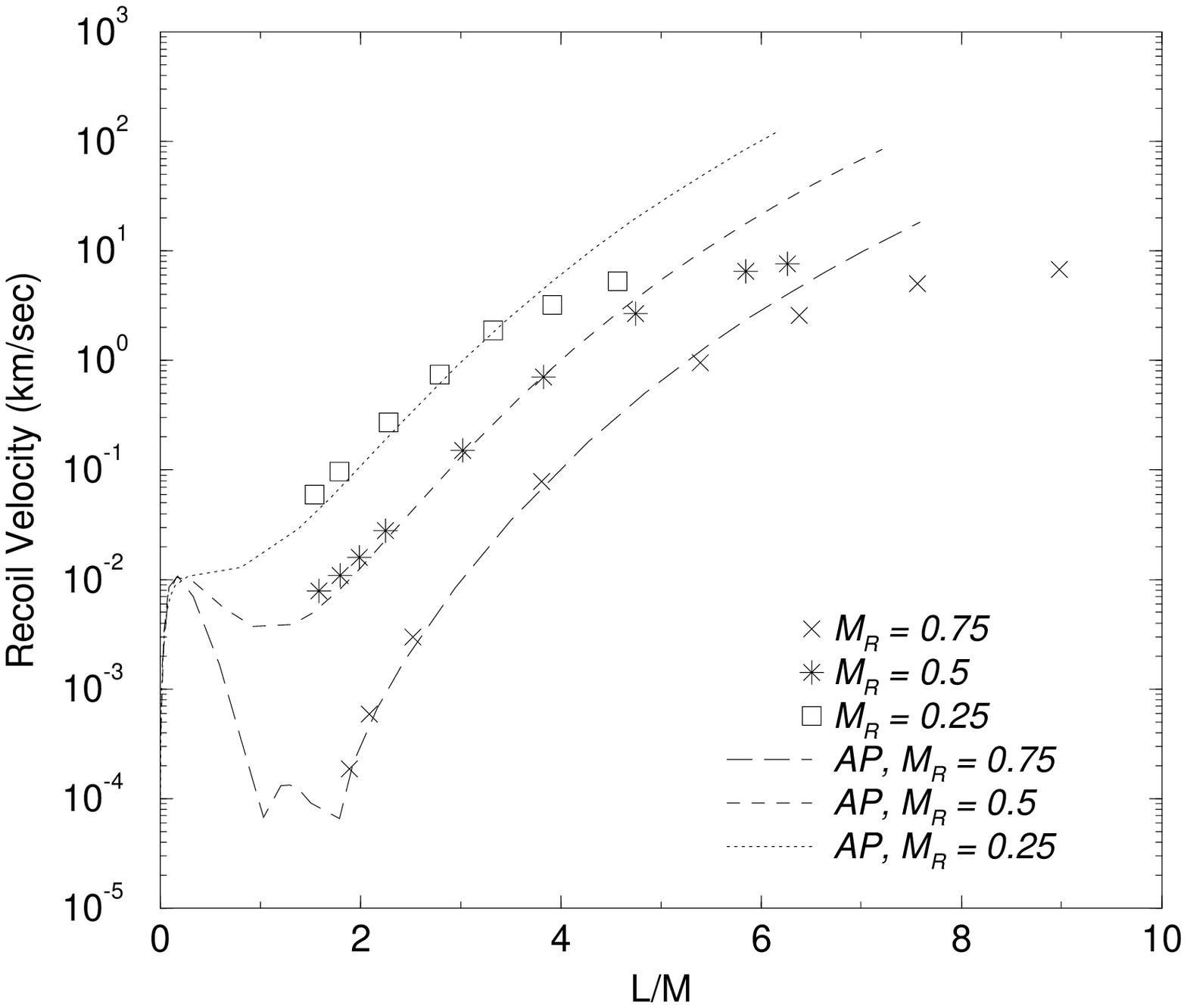}
\caption{
}
The absolute value of the recoil velocity is plotted together
with the CLAP results from Andrade and Price (AP) \cite{Andrade}
in units of km/sec for the three unequal mass
level curves ($M_R=0.75$, 0.5 and 0.25).
Due to the difficulty in resolving the weaker high--order multipole
components ($\ell \ge 3$) in the close separation limit, 
the numerical data is shown only for $(\mu_1, \mu_2) \simgt 0.5$, or
roughly $L/M \simgt 2$. The results here 
(and in Fig. \protect{\ref{fig:energy}}) are from simulations
run at a grid resolution of 
300$\times$70 (radial$\times$angular) zones.

\label{fig:momentum}
}

\end{document}